\documentclass[12pt]{iopart}

\usepackage{iopams}
\usepackage{graphicx,xspace}
\begin{document}

\title{Double percolation effects and fractal behavior in magnetic/superconducting
hybrids}

\author{L. Ruiz-Valdepe\~{n}as$^1$, M. V\'elez$^2$, F. Vald\'{e}s-Bango$^{2,3}$, L.
M. \'Alvarez-Prado$^{2,3}$, J.I. Mart{\'i}n$^{2,3}$, E.
Navarro$^{1}$, J. M. Alameda$^{2,3}$ and J. L. Vicent$^{1,4}$}
\address{$^1$Dpto. F{\'i}sica de Materiales, Universidad
Complutense, 28040 Madrid, Spain}
\address{$^2$Depto. F{\'i}sica, Universidad de
Oviedo, 33007 Oviedo, Spain}
\address{$^3$CINN, (CSIC - Universidad de Oviedo -
Principado de Asturias), Llanera, Spain}
\address{$^{4}$IMDEA-Nanociencia, Cantoblanco, 28049 Madrid, Spain}

\ead{mvelez@uniovi.es}

\begin{abstract}
 Perpendicular magnetic anisotropy ferromagnetic/ superconducting
 (FM/SC) bilayers with a labyrinth domain structure are used to study nucleation of
superconductivity on a fractal network, tunable through magnetic
history. As clusters of reversed domains appear in the FM layer,
the SC film shows a percolative behavior that depends on two
independent processes: the arrangement of initial reversed domains
and the fractal geometry of expanding clusters. For a full
labyrinth structure, the behavior of the upper critical field is
typical of confined superconductivity on a fractal network.

\end{abstract}

%\submitto{\NJP}

\maketitle
\section{INTRODUCTION}
Percolation phenomena are present in a wide variety of disordered
systems, ranging from random networks \cite{random} %to polymers \cite{polymer}
to nanomaterials \cite{nano}, granular superconductors
\cite{granular1,granular2} or perpendicular magnetic anisotropy
materials \cite{domain}. Theoretical understanding of these
phenomena has evolved from the classical analysis of simple
regular lattices \cite{lattice,alexander} to
more complex situations %that include continuum percolation and irregular lattices \cite{continumm} or many other different special lattices
\cite{bottleneck}. Recently, double percolation effects have been
theoretically introduced \cite{double} in order to describe
systems such as polymer blends \cite{blend} or nanomaterials
\cite{nano} in which disorder occurs on a two-level scale
characterized by two different geometrical length scales.

Hybrid ferromagnetic/superconducting (FM/SC) multilayers and
nanostructures are an interesting class of systems that are also
governed by the competition between two different length scales
\cite{reviewBuzdin}. The interplay between these two long range
phenomena results in a rich variety of behaviors such as reentrant
superconductivity \cite{reentrant}, domain wall superconductivity
(DWS) induced by the ferromagnetic exchange field
\cite{DWS0,DWS1,DWS2}, vortex guiding \cite{jaque,kwok,karapetrov}
or periodic vortex pinning \cite{review}. Magnetic domains have
been used to manipulate superconductivity both in parallel
\cite{DWS1,pissas} and perpendicular field configurations
\cite{gillijns,chien2}. In particular, stray fields created by a
magnetic layer with perpendicular magnetic anisotropy (PMA) have
proved a versatile tool to tune superconducting vortex pinning
\cite{alicia} and the superconducting phase diagram
\cite{chien,gillijns}, i.e. a controllable magnetic domain
structure can be created playing with the magnetic layer
hysteresis that, in turn, controls the nucleation of
superconductivity. Up to now, most attention has centered on
ordered domain geometries
 and on samples with relatively large
magnetic domains in comparison with the Ginzburg-Landau coherence
length $\xi_{GL}$ \cite{kwok,ordered,model1D_a,model1D_b}. In
these systems, due to field compensation effects, reentrant
superconductivity  is observed
\cite{reentrant,chien,ordered,model1D_a,model1D_b} and vortices
can be nucleated on top of the magnetic domain structure
\cite{iavarone,cieplak}. On the other hand, domains in PMA
materials often display a very disordered labyrinthine structure
\cite{domain,RETM} and peaks in the magnetoresistance curves of
PMA FM/SC multilayers have been reported as domain structure in
the FM layer changed from an ordered to a disordered configuration
indicating that the number and arrangement of domain walls can
have a significant influence on superconductivity \cite{chien2}.
Actually, percolation phenomena have been found in FM/SC bilayers
with a disordered domain structure in which, for large enough
domains, the superconducting sample fraction was directly given by
the reversed domains area \cite{yang}. However, taking into
account that labyrinthine domain structures can be described in
terms of a fractal geometry \cite{fractal1,fractal2}, they could
be used, for small enough domain sizes, to "design" a fractal
network for the nucleation of superconductivity similarly to
disordered superconducting wire networks (SWN) \cite{swn} and
granular superconductors \cite{granular1,granularAl}. Remarkably,
this labyrinthine structure would allow to tune a fractal
superconducting behavior through the magnetic film history.

In this work, we have studied PMA FM/SC bilayers with domain size
below 100 nm, that becomes comparable to $\xi_{GL}$ close to the
critical temperature $T_C$. We show that nucleation of
superconductivity is controlled both by the distribution of
clusters of reversed domains and their fractal geometry resulting
in a double level percolation process. Once the percolation
process is finished and the labyrinthine domain configuration
extends homogenously through the sample, the upper critical field
$H_{c2}$ shows the characteristic temperature dependence of
confined superconductivity on a fractal network.

\section{EXPERIMENTAL}

FM/SC NdCo/Nb bilayers have been fabricated by sputtering on 1 cm
$\times$ 1 cm Si(100) substrates in a two step process. First, a
NdCo$_5$($t$ nm) amorphous layer is grown by cosputtering
\cite{hierro,cid} with thickness ($t$) in the 40 - 80 nm range on
a Si substrate covered by a 10 nm thick Al buffer layer and, then,
a 5 nm thick Al capping layer is grown on top. Next, the
Al/NdCo/Al sample is taken out of the chamber so that the Al
capping layer becomes oxidized and, finally, a Nb film with
thickness $d_{Nb}$ = 50 nm is grown on top by sputtering to get
the complete FM/SC bilayer structure.

The properties of the SC layer have been characterized on a
control 50 nm Nb film, grown on a bare Si substrate under similar
conditions. It presents a $T_C$ = 7.55 K, typical for Nb in this
thickness range \cite{minhaj} and a Ginzburg-Landau coherence
length $\xi_{GL}(0) = 8.95$ nm, obtained from the temperature
dependence of $H_{c2}$ = $\Phi_0/2\pi \xi_{GL}(T)^2$ with $\Phi_0$
the quantum of flux. It corresponds to a superconducting coherence
length $\xi_s = \frac{2}{\pi}\xi_{GL}(0) = 5.7$ nm
\cite{radovic,alija}.

The FM layer is made of NdCo$_5$, an amorphous alloy with
saturation magnetization $M_S \approx 10^3$ emu/cm$^3$. It
presents a moderate room temperature (RT) perpendicular magnetic
anisotropy, that can be characterized by an out-of-plane uniaxial
anisotropy constant \cite{libro} ($K_n$), with values of the order
$K_n \approx 10^6$ erg/cm$^3$. Upon lowering the temperature,
$K_n$ increases up to $K_n \approx 10^7$ erg/cm$^3$ at 10 K
\cite{cid}, that is, the anisotropy ratio $Q = K_n/2\pi M_S^2$ is
$Q \approx 0.1$ at room temperature and $Q \approx 1$ at 10 K.

In general, proximity effects between the FM and SC layers could
be caused both by exchange and stray fields. In the present case,
for $Q \approx 1$, the domain structure in the PMA Nd-Co layer
should contain significant out-of-plane magnetization components
that can be used to create relatively large stray fields in the
neighbouring Nb layer. On the other hand, the oxidized AlO$_x$
layer in between Nb and NdCo layers should act as an
exchange-field insulator so that proximity effects due to the
exchange field in the FM layer should be small. Also, $d_{Nb}
>> \xi_s$ in these films implying that exchange induced DWS should
be strongly suppressed \cite{DWS2}. In any case, the aim of our
study is mainly on the effect of geometry on the superconducting
regions that nucleate on top of the magnetic domains rather than
on the specific proximity effect mechanism (exchange field and/or
stray field) involved to create them.

Magnetic and superconducting properties of the SC/FM bilayers have
been studied by magnetotransport measurements in a He cryostat
equipped with a 90 kOe superconducting solenoid and a rotatable
sample holder that allows to vary \textit{in-situ} the applied
field direction from in-plane to out-of-plane. Transport
measurements are performed on extended samples using a four probe
DC technique in a van der Pauw
configuration\cite{vanderPauw,cagigal} which is often used in
percolation problems \cite{schad,schad2,schad3} with an applied
current ($I_{DC} = 10-100 \mu$A). Depending on whether voltage and
current contacts are consecutive or crossed along the sample edge
either the resistance $R$ or Hall effect $R_{Hall}$ signals can be
obtained from exactly the same sample area \cite{vanderPauw}.

First, the magnetic properties of the FM layer have been obtained
from Hall effect measurements making use of the much larger Hall
effect in ferromagnetic materials \cite{EHE} than in ordinary
metals such as Nb. Briefly, in a FM layer with an out-of-plane
magnetization component $M_z$ under a perpendicular field $H_z$,
the Hall signal ($R_{Hall}$) is given by $R_{Hall} = R_0H_z +
R_{EHE}M_z$, with $R_0$ the ordinary Hall effect coefficient,
related with the deviation of charge carriers by the Lorentz
force, and $R_{EHE}$ the Extraordinary Hall effect (EHE)
coefficient, related with spin dependent scattering of conduction
electrons \cite{EHE}. $R_{EHE}$ is usually much larger than $R_0$
(about a factor 10-100) and is particularly enhanced in Rare
Earth-Transition Metal amorphous alloys due to their large
resistivities and strong spin-orbit coupling \cite{EHE2}. Thus,
Hall effect hysteresis loops of the FM/SC bilayers should be
dominated by the EHE term $R_{EHE}M_z$ in the Nd-Co layer
\cite{EHE2,valdes}. This procedure allows us to characterize the
FM layer magnetic properties at 10 K, just above $T_c$, and, also,
to control  its magnetic history \emph{in situ} by performing
different kinds of minor hysteresis loops. Then, once the sample
has been prepared in the desired magnetic state at 10K, its
superconducting properties have been characterized measuring
resistance transitions ($R(T)$ curves) as the temperature is
lowered under a constant out-of-plane magnetic field $H_z$.

\section{RESULTS AND DISCUSSION}
\subsection{Magnetic properties and stray field of the NdCo FM layers}

 Hysteresis loops
in PMA films can have a variety of shapes depending on the
relative strength of PMA, exchange and disorder \cite{jagla}.
\begin{figure}[ht]
 \begin{center}
\includegraphics[angle=0,width=1\linewidth]{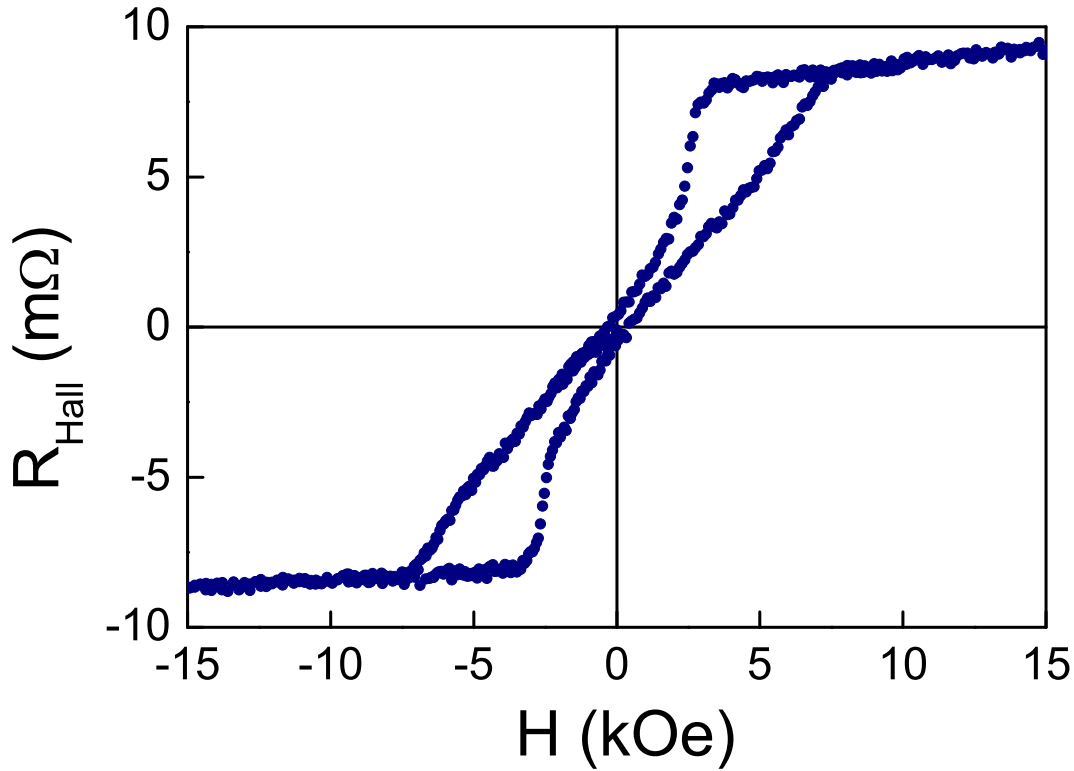}

 \caption{(color online) EHE hysteresis loop of a 52 nm NdCo/50 nm Nb bilayer at 10K. Note that a constant resistance offset $R_{offset}=1.2 \Omega$
 due to contact misalignment has been subtracted to obtain $R_{Hall}$ from raw resistance measurements.} \label{ciclo}
 \end{center}
 \end{figure}
Figure 1 shows the EHE out-of-plane loop for a 52 nm NdCo/Nb
sample measured at 10 K, i.e. just above $T_C$. It presents the
characteristic shape of magnetization reversal through the
nucleation and expansion of clusters of stripe domains
\cite{jagla,ciclo}: upon decreasing $H_z$ from saturation,
$R_{Hall}$ follows a weak linear field dependence due to ordinary
Hall effect while small reversed domains begin to nucleate across
the sample; then, there is a steep decrease in $R_{Hall}$ already
at positive fields, between 3.4 kOe $> H_z
>$ 2.1 kOe, that marks the sudden growth of clusters of labyrinth
stripe domains until they fill the whole sample area; next, there
is an almost reversible regime in which $R_{Hall}$ decreases
linearly with $H$ due to relative changes in the width of "up" and
"down" domains and the film displays a very low remanence and
coercivity. Finally, as domains with the initial magnetization are
annihilated, negative saturation is reached. Upon increasing the
temperature up to RT, the decrease in PMA in the Nd-Co films
results in an enhancement of the low field reversible region while
high field hysteresis almost disappears \cite{jpd}. Both the
remanent magnetization ($M_R$) and coercivity ($H_{coer}$) stay
almost constant at very low values in the whole temperature range
from 10K to RT ($M_R$ below $0.05M_S$ and $H_{coer}$ below 0.5
kOe) indicating that there should not be qualitative changes
between the room temperature and the 10 K remanent domain
structure.

Figures 2(a) and (b) show the differences in the remanent domain
configurations of NdCo films depending on their previous magnetic
history, obtained by Magnetic Force Microscopy (MFM) at room
temperature \cite{hierro}: in the first case (figure 2(a)), an
in-plane field $H_y = 1$ kOe has been applied to a 45 nm thick
NdCo sample and, then, it has been decreased to zero creating the
typical parallel stripe domain structure oriented along $H_y$ with
stripe domain period $\Lambda = 115$ nm; in the second case
(figure 2(b)), a labyrinthine domain configuration is observed in
an 80 nm thick NdCo film after saturation with an out-of-plane
$H_z = 4$ kOe. $\Lambda$ values obtained from the RT MFM
characterization depend on $t$ and $H$, but they are in the 100 nm
- 300 nm range for the studied samples \cite{hierro,valdes}.

\begin{figure}[ht]
\begin{center}
\includegraphics[angle=0,width=0.7\linewidth]{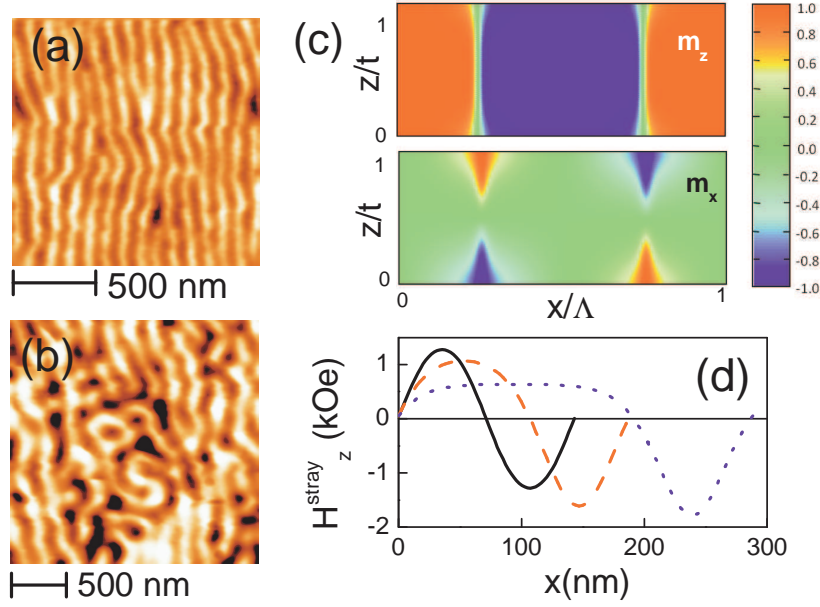}

 \caption{(color online) (a) MFM image of a 45 nm
  NdCo film at remanence after applying an in-plane $H_y = 1$ kOe ($\Lambda$ = 115 nm); (b) MFM image of a 80 nm
  NdCo film at remanence after applying an out-of-plane $H_z = 4$
kOe;(c) Map of the simulated magnetization distribution
  of a 52 nm NdCo film at remanence: top panel, out-of-plane $m_z$ and bottom panel, in-plane $m_x$. Note the reduced scale for
  spatial variables, $z/t$ and $x/\Lambda$ ($t=52$ nm and $\Lambda = 143$ nm). Color code of $m_x$ and $m_z$ values is indicated in left scale bar.
  (d) Simulated $H^{stray}_{z}(x)$ created by
  a 52 nm NdCo film at $h$ = 30 nm plotted over one stripe period $\Lambda$ for:  $H_z = 0$, solid line; $H_z = 3$ kOe, dashed line; $H_z = 4.5$ kOe, dotted
  line.
  }
  \end{center}
  \label{Fabricacion}
 \end{figure}

The effect of the temperature dependence of $K_n$ on the low
temperature domain structure has been studied by micromagnetic
simulations \cite{micromagnetism}. First, material parameters have
been adjusted to reproduce the observed RT domain structure and,
then, the temperature variations of $K_n$ and $M_S$ have been
introduced in order to calculate the low temperature equilibrium
parallel stripes domain configuration, that is quite similar to RT
but with a small increment in $\Lambda$ (about 10\%). Figure 2(c)
shows a map of the magnetization distribution (out-of-plane
$m_z=M_z/M_s$ and in-plane $m_x=M_x/M_s)$ obtained for a 52 nm
NdCo film at 10 K calculated with anisotropy constant $K_n =
5.6\times 10^6$ erg/cm$^3$, magnetization $M_S = 897$ emu/cm$^3$,
and exchange $A = 10^6$ erg/cm at $H_z=0$. Top panel in figure
2(c) shows the typical alternating "up"/"down" domains in $m_z$
separated by Bloch walls that make up the periodic parallel stripe
configuration with $\Lambda = 143$ nm. Domain wall width
$\delta_w$ at the film surface can be estimated from the size of
the region in which $m_z$ turns from 1 to -1. It is about 40 nm in
the simulations with RT parameters and goes down to 18 nm at low
temperature. Bottom panel in figure 2(c) allows us to observe the
existence of Neel caps on top of the Bloch walls in which $M$
becomes parallel to the film surface. Thus, due to the moderate
PMA of these films, the domain structure is quite different from
previously studied PMA FM/SC systems ($Q>>1$) \cite{model1D_b}. It
not only contains parallel regions with oscillating out-of-plane
magnetization component but, as well, a relatively large
flux-closure structure of Neel caps close to the film surface that
significantly weakens the stray field.

Now, we can calculate the stray field $H^{stray}_z$ created by the
simulated parallel stripe domain structure in the space above the
FM layer. figure 2(d) is a plot of the stray field profile at the
mid-plane of the Nb film (at a height $h=30$ nm) starting from the
equilibrium remanent state ($H=0$) and, then, upon applying an
out-of-plane field $H_z$ of increasing magnitude. At remanence,
$H^{stray}_z$ displays a symmetric profile with period $\Lambda =
143$ nm. Then, as $H_z$ increases, the stray field profile becomes
asymmetric due to the growth of positive domains at the expense of
the negative ones and $\Lambda$ becomes larger going up to 288 nm
at 4.5 kOe. At the same time, as the size of negative domains
shrinks, the average negative $\langle H^{stray}_z \rangle$,
calculated as the spatial average of $H^{stray}_z$ over the
negative domain region, is enhanced from -825 Oe at remanence to
-1 kOe at $H_z = 4.5$ kOe.

\begin{figure}[ht]
\begin{center}
\includegraphics[angle=0,width=1\linewidth]{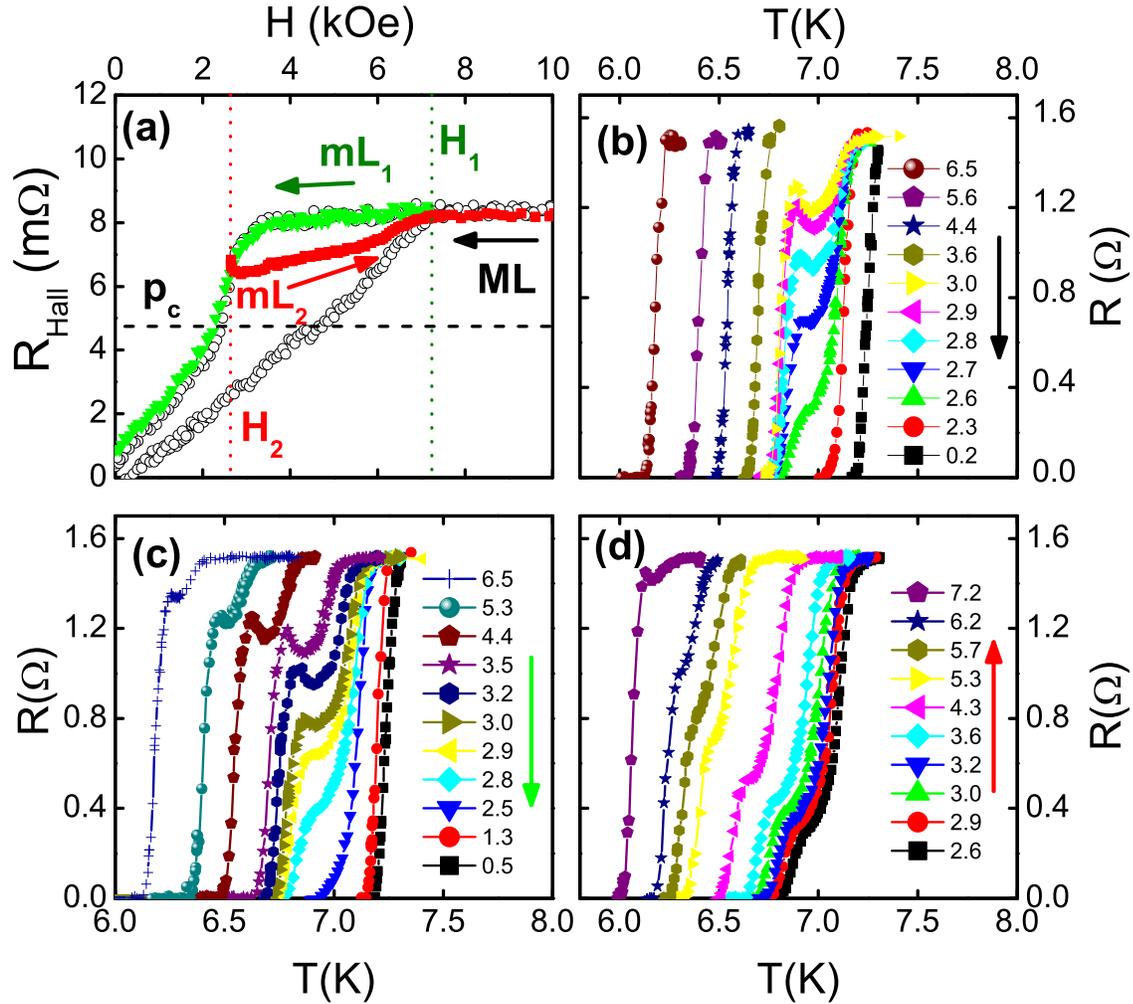}

 \caption{(color online)(a) EHE loops  of a 52 nm NdCo/50 nm Nb bilayer at 10K along three hysteresis processes:
 $\circ$, Major loop (ML), measured as $H_z$ decreases from saturation at 20 kOe;
  $\blacktriangledown$, minor loop 1 (mL$_1$), measured as $H_z$ decreases from incomplete saturation at $H_1 = 7.2$ kOe;
  $\blacksquare$, minor loop 2 (mL$_2$), $H_z$ goes down from 20 kOe to $H_2$ = 2.6 kOe and, then, $R_{Hall}$ is measured upon increasing
  $H_z$. Dashed line indicates the signal level at 2.45 kOe (i.e. at the percolation threshold in ML), used to estimate $p_c$ as indicated in the text.
  Sets of consecutive $R(T)$ curves at constant $H_z$, measured along (b)
  ML; (c) mL$_1$ and (d) mL$_2$. Labels indicate $H_z$ in kOe at each $R(T)$.} \label{transiciones}
 \end{center}
 \end{figure}

 \subsection{Superconducting transitions and percolation effects}

 Figure 3 shows the superconducting transitions measured on the 52
nm NdCo/Nb bilayer at a constant $H_z$, following different
hysteresis paths in the magnetic layer as indicated in figure
3(a). One of them is a major hysteresis loop (ML, $\circ$): a
large $H_z = 20$ kOe is applied to the sample in order to reach an
out-of-plane saturated state and, then, R(T) curves are measured
at decreasing $H_z$ values. The other two processes are minor
hysteresis loops. In the first minor loop (mL$_1$,
$\blacktriangledown$), $H_z$ starts at -20 kOe, i.e. with the
sample at the negative out-of-plane saturated state, then it
increases up to $H_1 = 7.2$ kOe which is the field at which
irreversibility disappears in the loop but is not large enough to
ensure full out-of-plane saturation (i.e. small inverted domains
could still persist in the sample with a very small contribution
to the magnetization \cite{ciclo}) and, finally, R(T) curves are
measured as $H_z$ decreases from $H_1$. In the second minor loop
(mL$_2$, $\blacksquare$) $H_z$ starts at 20 kOe, i.e. with the
sample at the positive out-of-plane saturated state, then $H_z$ is
taken down to $H_2$ =2.6 kOe in order to bring the sample to the
initial stages of reversed domain expansion and, then, R(T) curves
are measured  as $H_z$ increases so that the reversed area
fraction must shrink again. In all the cases there is a certain
field range in which the superconducting transitions develop a
two-step structure, characteristic of the break up of the sample
in two kinds of regions with different $T_C$'s ($T_{C1}$ and
$T_{C2}$). At intermediate temperatures, $T_{C1}<T<T_{C2}$,
conduction takes place by percolation through the network of
coexisting superconducting and normal regions within the sample
\cite{yang,chiodi}. It is interesting to note that the field range
of occurrence of this two-step percolative behavior is completely
different between the major loop (3 kOe $\geq H_z \geq 2.5$ kOe)
and mL$_1$ (6.5 kOe $\geq H_z \geq$ 2.8 kOe) in spite of the
almost identical $R_{Hall}(H_z)$ curves. This can be directly
attributed to the small reversed domains present in the FM layer
due to incomplete saturation during mL$_1$ loop that serve as
nucleation centers for superconductivity in a wider field range.

A more detailed analysis of the two step transitions is shown in
figure 4 in which the temperature dependence of the upper critical
field $H_{c2}(T)$ is plotted. The phase boundary for
superconductivity is usually obtained from $R(T)$ curves
\cite{reentrant,chien,karapetrov,tinkham,vicent} as the points in
the $H-T$ plane in which $R$ is a certain fraction of the normal
state resistance $R_n$.
 In this case, we have
used two different resistance criteria $0.1R_n$ (to obtain
$H_{c2}^{0.1R_n}(T)$) and $0.9R_n$ (to get $H_{c2}^{0.9R_n}(T)$),
in order to characterize the two different kinds of regions
present in the sample. For the major loop ML (figure 4(a)), both
curves run essentially parallel at high and low fields at 0.1 K
distance. This is similar to the transition width of plain
reference Nb films indicating that there is essentially no field
broadening of the superconducting transitions whenever the FM/SC
bilayer is in a homogeneous state. However, at intermediate fields
they become clearly separated as steps develop in the $R(T)$
curves: below 3 kOe, there is a sudden jump in
$H_{c2}^{0.9R_n}(T)$ from the typical linear dependence ($H_{c2}=
41.2$ kOe$(1-T/T_C)$) to a different non-linear temperature
dependence that could be an indication of confined
superconductivity \cite{alexander,granular1}; $H_{c2}^{0.1R_n}(T)$
follows a similar trend but it retains the linear temperature
dependence down to 2.5 kOe. That is, percolation effects appear
approximately in the field range corresponding to the
nucleation/expansion of clusters of reversed domains (shaded areas
in figure 4). The different temperature dependence of
$H_{c2}^{0.9R_n}$ and $H_{c2}^{0.1R_n}$ for 3 kOe $> H > 2.45$ kOe
is a signature of qualitative differences between the coexisting
normal/superconducting regions during the percolation process:
$H_{c2}^{0.9R_n}$ corresponds to confined superconducting regions
that nucleate on top of clusters of reversed domains (see sketch
in figure 4(a)) whereas $H_{c2}^{0.1R_n}$ marks the transition of
the surrounding extended areas that retain the same dimensionality
as the continuous film. Then, for $H <$ 2.45 kOe, steps in
$R(T)$'s disappear as the confined superconducting areas percolate
through the sample effectively shorting the possible remaining
extended normal regions. The reversed area fraction $p_{mag}$ at
each point of the hysteresis loop can be estimated as $p_{mag} =
0.5(1-M/M_S)$. Thus, the percolation threshold $p_c$ would
correspond to $p_{mag}$ at 2.45 kOe, i.e. $p_c \approx 0.2$.
\begin{figure}[ht]
\begin{center}
\includegraphics[angle=0,width=1\linewidth]{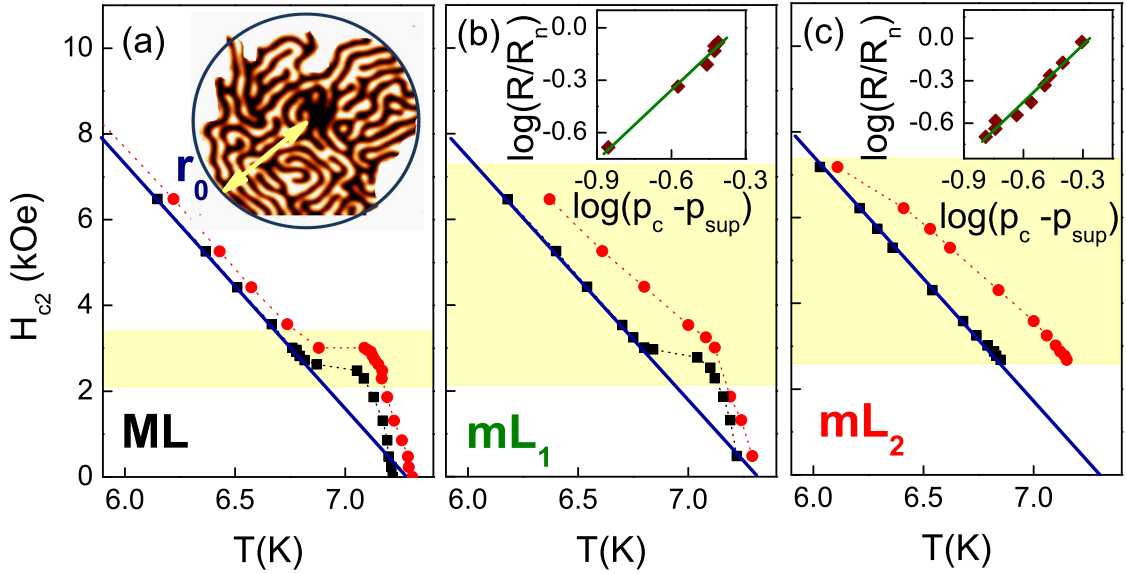}

 \caption{(color online)$H_{c2}$ \textit{vs.} $T$ defined at 0.1$R_n$ ($\blacksquare$) and at 0.9$R_n$ ($\bullet$)
 measured along (a) ML (b) mL$_1$ and (c) mL$_2$. Solid lines are linear fits to $H_{c2}= 41.2$ kOe$(1-T/T_C)$.
 Shaded areas mark field range of nucleation/expansion of inverted domains obtained from
EHE loops. Inset in (a) is a sketch of a cluster of inverted
domains. Insets in (b) and (c) are log-log plots of $R/R_n$ at the
plateaus in $R(T)$ \textit{vs.} $(p_c-p_{sup})$.}
\end{center}
\label{percolation}
 \end{figure}

The qualitative differences between $H_{c2}^{0.9R_n}(T)$ and
$H_{c2}^{0.1R_n}(T)$ are more evident in figures 4 (b) and (c)
corresponding to minor loops mL$_1$ and mL$_2$. Now, due to FM
film history, reversed domains and, therefore, steps in $R(T)$
curves are present in a wider field range, which allows for a more
thorough analysis of the percolation process. According to
classical percolation theory in 2D \cite{percolacion}, the
resistance of a random mixture of superconducting/normal elements
scales with the superconducting fraction $p_{sup}$ as
\begin{equation}
R \propto (p_c-p_{sup})^{s}
\end{equation}
with universal exponent $s = 1.3$. The percolation threshold is
$p_c = 0.5$ for square lattice models, but it can vary
significantly depending on system geometry \cite{disks}. However,
the resistance values at the "plateaus" in $R(T)$ curves in figure
3 cannot be described by eq. (1) using the simplest assumption
$p_{sup} = p_{mag}$ that was used in Nb/BaFeO hybrids with much
larger domain sizes \cite{yang}. This is reasonable since, the
comparison between ML and mL$_1$ data directly shows that
effective superconducting area is not simply proportional to the
inverted domain area but depends on the previous magnetic history
(i.e. on the initial distribution of reversed domains). Also, the
small observed $p_c\approx 0.2$ is typical of two-level
percolation \cite{double,blend} in which two independent random
processes are at play (reversed domain nucleation and propagation
in the PMA film here \cite{RETM}). In the first level, we may
consider a random arrangement of initial reversed nuclei and, on
the second level with a finer length scale, the fractal expansion
of each cluster of reversed domains starting from each initial
nucleus \cite{RETM,fractal1,fractal2}. Thus, along the
superconducting transition, the Nb film will be composed of a
random array of SC islands nucleated on each of these clusters of
reversed domains. For a labyrinth domain structure, we may
consider that the effectively shorted area for each SC island is
that of a disk enclosing the cluster \cite{coniglio} (see sketch
in figure 4(a)). The radius of this circle $r_0$ scales as
$p_{mag} \propto r_0^{D_m}$ with $D_m$ the mass dimension of the
cluster \cite{RETM} ($D_m = 1.896$ for an infinite cluster
\cite{disks} and $D_m\approx 1.5-2$ reported for expanding
clusters of reversed domains in amorphous Rare Earth-Transition
Metal alloys \cite{RETM}). Thus, the area of SC islands should
scale as $p_{sup} \propto r_0^2 \propto (p_{mag})^{2/D_m}$. Insets
in figures 4 (b) and (c) show the results of the fit of $R$
\emph{vs.} $p_{sup} = (p_{mag})^{2/D_m}$ to equation (1). A linear
behavior in the log-log plot appears for $D_m=1.8$, close to the
mass dimension of an infinite cluster, and $s = 1.32 \pm 0.07$, in
good agreement with the critical exponent expected from 2D
percolation theory \cite{percolacion}. $s$ and $D_m$ are similar
both for mL$_1$ and mL$_2$ but certain differences appear in the
fitted thresholds $p_c($ml$_1$) = 0.16 and $p_c($ml$_2$) = 0.21,
which may be attributed to the different distribution of initial
reversed nuclei in each hysteresis process.

\subsection{Upper critical field dimensionality}

Further confirmation on the fractal geometry of the
superconductivity phase nucleated on top of the labyrinth reversed
domains may be obtained from the analysis of the temperature
dependence of $H_{c2}$, focusing in the homogeneous regime after
the percolation process is finished and steps have disappeared
from the $R(T)$ transitions. Figure 5 shows several
$H_{c2}^{0.1R_n}(T)$ lines measured along different hysteresis
processes. First, there are two out-of-plane major loops both in
the descending and ascending field branches: ML1(-) with $H_z$
decreasing from saturation at 20 kOe, ML1(+) with $H_z$ increasing
from remanence after saturation at -20 kOe, ML2(-) with $H_z$
decreasing from saturation at 90 kOe, ML2(+) with $H_z$ increasing
from remanence after saturation at -90 kOe. Data are also included
upon increasing $H_z$ from different remanent states: Demag1(+)
and Demag2(+) correspond to out-of-plane demagnetized states after
performing a series of $H_z$ cicles of decreasing amplitude in
order to create a disordered labyrinth domain pattern over the
whole sample and Ordered(+) corresponds to an in-plane remanent
state after applying an in-plane $H_y = 90$ kOe in order to create
an ordered parallel stripes configuration perpendicular to the
applied current direction. In the two descending field processes
ML1(-) and ML2(-), steps in the R(T) transition disappear below
2.5 kOe, as discussed above, indicating that the lower field
region corresponds to a fully percolated labyrinthine state. On
the other hand, in all the other ascending field processes no
steps have been observed in the R(T) transitions indicating that
their initial remanent state covers the sample homogenously. Thus,
the data in figure 5 will allow us to study the characteristic
dimensionality of the superconducting state nucleated on top of
different homogeneous domain structures.
\begin{figure}[ht]
\begin{center}
\includegraphics[angle=0,width=1\linewidth]{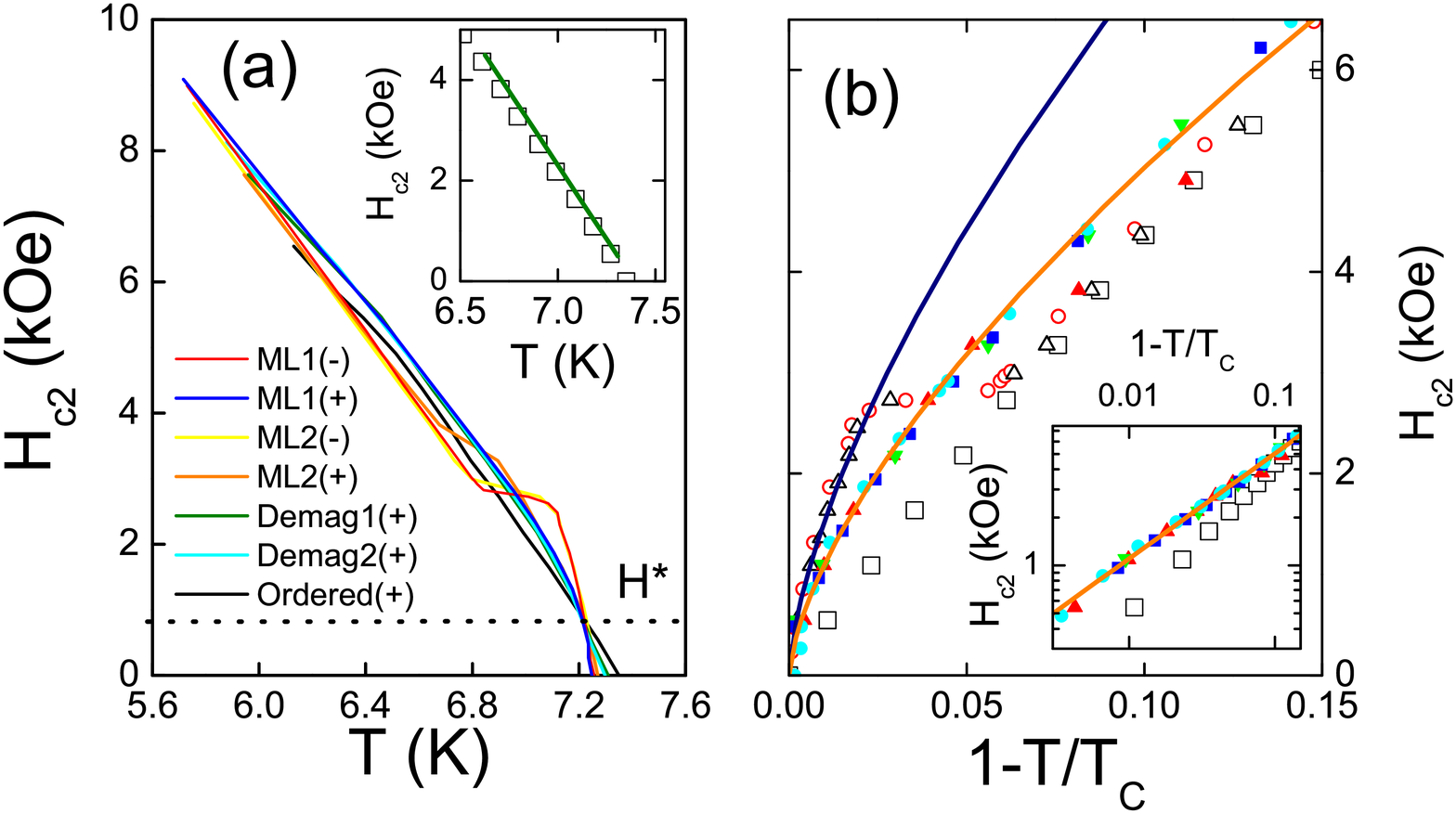}

 \caption{(color online)(a) $H_{c2}^{0.1R_n}(T)$ measured following different hysteresis
  process in the magnetic layers. Note the crossover at $H^*=0.8 kOe$.
  Inset shows $H_{c2}^{0.1R_n}(T)$ upon increasing $H_z$ after in-plane
  saturation($\square$, Ordered(+))
  in comparison with the simulated dependence with a 1D model for a FM/SC bilayer with $\langle H_z^{stray}(h = 55$
  nm$)\rangle$ (solid line);
  (b) $H_{c2}^{0.1R_n}$ \emph{vs.} $1-T/T_C$. Parallel stripes remanent state: $\square$, Ordered(+); labyrinthine states: $\circ$, ML1(-);
   $\bigtriangleup$, ML2(-);
   $\blacktriangledown$, Demag1(+); $\blacksquare$, Demag2(+); $\bullet$, ML1(+); $\blacksquare$
   ML2(+). Solid lines are fits to $H_{c2} \propto (1-T/T_C)^n$ with $n=0.66$.
   Inset shows same temperature dependence in a log-log scale.} \label{fractal}
 \end{center}
 \end{figure}
It can be seen that these $H_{c2}^{0.1R_n}(T)$ curves are strongly
dependent on magnetic history but present a common crossover at
$H_z^* = 0.8 $ kOe: the processes with stronger $H_{c2}$
enhancement at high fields present also the lower $T_C(H=0)$.
However, in the reduced temperature scale $1-T/T_C$,
$H_{c2}^{0.1R_n}$ data become grouped in three distinct sets (see
figure 5 (b)) depending on stripe domain geometry: one of them
(Ordered(+)) corresponds to $H_z$ increasing from remanence after
in-plane saturation, i.e. from an ordered parallel stripe remanent
state, and the other two correspond to labyrinth domain
configurations either as $H_z$ decreases below the percolation
threshold (ML1(-) and ML2(-)) or as $H_z$ increases from remanence
after out-of-plane negative saturation or demagnetization (ML1(+),
ML2(+), Demag1(+) and Demag2(+)).

$H_{c2}(T)$ in the parallel stripes case can be analyzed in terms
of existing stray field induced superconductivity models for SC/FM
bilayers with an ordered lateral geometry
\cite{model1D_a,model1D_b}. In particular, we have used a 1D model
in the small domain size limit \cite{model1D_b} complemented with
the stray field values obtained from the micromagnetic
calculations. Nucleation of superconductivity is determined by
field confinement effects within a length scale $L_H$ so that the
upper critical field is given by the condition $\xi_{GL}(T)
\approx L_H$. Taking into account the superposition of the stray
field created by the FM layer $H_z^{stray}$ with the applied field
$H_z$, $L_H$ is given by \cite{model1D_b}
\begin{equation}
L_H = (\Phi_0/2\pi|H_z-H_z^{stray}|)^{1/2}. \end{equation}
 In our case, $H_z^{stray}$ is a function of $H_z$, as shown in figure 2(d), so that equation (2)
 allows to estimate $H_{c2}$ through the implicit condition
 $H_{c2}(T) = H_z^{stray}(H_{c2}) + \Phi_0/2\pi\xi_{GL}(T)^2$. Best fit is obtained
taking $H_z^{stray}(H_z)$ from the micromagnetic calculations at
the top surface of the Nb film, $H_z^{stray} = \langle
H_z^{stray}(h=55 $nm$)\rangle$ (see solid line in the inset of
figure 5(a)). This is reasonable since nucleation of
superconductivity should be favored at the top SC film surface in
which the superconducting order parameter is maximum
\cite{model1D_b}. At low fields, $L_H$ increases and, eventually,
becomes larger than domain size, so that the superconducting wave
function extends over several domains and any $H_{c2}$ enhancement
due to the stray field disappears since it is averaged over
positive and negative domains \cite{model1D_b}. There is still a
certain $T_C(H=0)$ reduction due to the inhomogeneity introduced
in the system by the periodic domain structure \cite{model1D_a}.
At the crossover field found in figure 5(a), $H_z^*$ = 0.8 kOe,
$\xi_{GL} =
 63$ nm and $L_H^* = 75 $nm, which are comparable to domain size $\Lambda/2 =
71.5$ nm, indicating that it can correspond to the crossover from
extended to localized superconductivity: more disordered domain
structures with a stronger stray field present a larger $H_{c2}$
enhancement in the high field range of localized superconductivity
over reversed domains but, also, a more important $T_C(H=0)$
reduction in the low field extended superconductivity regime.

Finally, let us focus on the two sets of $H_{c2}^{0.1R_n}$ data in
figure 5(b) measured with labyrinth domain configurations that
present a different behavior from the simple 1D ordered parallel
stripe geometry. In spite of the differences in magnetic history
(either out-of-plane saturated or out-of-plane demagnetized
initial state), all these data follow clearly a stronger
temperature dependence $H_{c2} \propto (1-T/T_C)^n$ than expected
for a 2D superconducting film in a perpendicular field geometry.
In both cases the same $H_{c2}$ exponent $n=0.66$ is found,
indicating that it is an intrinsic property of the labyrinth
domain geometry. It lies in between 2D and 1D values $n = 1$ and
0.5, respectively, and is similar to reported values in granular
superconductors \cite{granular1,granularAl} and disordered SWN
\cite{swn}. This reduced $H_{c2}$ exponent has been attributed to
the fractal nature of the percolation networks \cite{alexander},
with $n=0.69$ predicted for the infinite cluster, in good
agreement with the experimental results in figure 5.

\section{CONCLUSIONS}

In summary, the disordered labyrinth domain structure of PMA NdCo
layers have been used to "imprint" a fractal geometry in the
superconducting state of NdCo/Nb bilayers. Superconducting
transitions display a characteristic percolative behavior with the
Nb film broken up into a mixture of extended normal regions and
islands of confined superconductivity. The distribution and size
of these islands is controlled by a two level percolation process:
first, by the random distribution of initial reversed nuclei
determined by magnetic film history and, second, by the fractal
expansion of each cluster of reversed domains resulting in a
larger effective superconducting area. The dimensionality of
$H_{c2}(T)$ lines can also be tuned by magnetic film history: when
it adopts a parallel stripe domain configuration, $H_{c2}(T)$ can
be described by a simple model of field confinement with a 1D
domain pattern; however, when domains adopt a labyrinth geometry,
$H_{c2}$ displays a stronger temperature dependence $H_{c2}
\propto (1-T/T_C)^{0.66}$ characteristic of fractal
superconducting networks.

\ack
 Work supported by Spanish MINECO under grants FIS2008-06249,
Consolider CDS2007-00010 and CAM grant S2009/MAT-1726. LRV
acknowledges support from a FPU fellowship.

\end{document}